\def\lengthOfSequence{L}
\def\WUStime[#1][#2]{x_{#1}(#2)}
\def\signalTime[#1][#2]{f_{#1}(#2)}
\def\signalTimeConjugated[#1][#2]{f^*_{#1}(#2)}
\def\amplitudeTime[#1][#2]{a_{#1}\left(#2\right)}
\def\phaseTime[#1][#2]{\theta_{#1}\left(#2\right)}
\def\amplitudeShape[#1]{{\bf a}_{#1}}
\def\phaseShape[#1]{{\boldsymbol \uptheta}_{#1}}

\def\lagrangeVariable{\lambda}

\def\Tactive{T_{\rm active}}
\def\indexIteration{i}

\def\offSetDuration[#1]{\delta_{#1}}

\def\elementOfAmplitudeShape[#1][#2]{{\{{\bf a}_{#1}\}}_{#2}}
\def\elementOfPhaseShape[#1][#2]{{\{{\boldsymbol \uptheta}_{#1}\}}_{#2}}

\def\DFTmatrix[#1]{{\bf F}}
\def\mappingMatrix{{\bf M}}
\def\amplitudeMatrix{{\bf A}}

\def\shapeTime[#1][#2]{\Lambda_{#1}\left(#2\right)}

\def\cpDuration{T_{\rm CP}}
\def\symbolDuration{T_{\rm s}}
\def\expectedValue[#1]{\mathbb{E}\left\{#1\right\}}

\def\diag[#1]{{\rm diag}\left\{#1\right\}}
\def\deltaF{\Delta f}
\def\bit[#1]{b_{#1}}

\def\sequenceOptimal[#1]{\hat{\bf s}_{#1}}
\def\sequenceOptimalIterations{{\bf \bar{s}}}
\def\sequence[#1]{{{\bf s}_{#1}}}
\def\sequenceFlipped[#1]{{{\bf \tilde{s}}_{#1}}}
\def\sequenceConjugated[#1]{{{\bf s}_{#1}^*}}

\def\elementOfSequenceMapped[#1][#2]{{\{{\bf s}_{#1}\}}_{#2}}
\def\waveformBasis{\mathcal{W}}
\def\numberOfTaps{L}
\def\power[#1]{P_{#1}}

\def\boundLeakage{u_{\rm leak}}
\def\boundFirst{u_{\rm first}}
\def\powerFirstSample{P_{\rm first}}
\def\powerLeakage{P_{\rm leak}}

\def\onPeriod[#1]{\mathbb{M}_{{\rm high}#1}}
\def\offPeriod[#1]{\mathbb{M}_{{\rm low}#1}}

\def\onPeriodSampled[#1]{\mathbb{N}_{{\rm high}#1}}
\def\offPeriodSampled[#1]{\mathbb{N}_{{\rm low}#1}}

\def\numberOfSamples{N}
\def\indexSample{p}

\def\cost[#1]{\epsilon_{#1}}
\def\costSampled[#1]{\epsilon_{{\rm s}#1}}
\def\costSampledModified[#1]{\epsilon_{{\rm m}#1}}

\def\indexSequence{i}
\def\timeVariable{t}

\def\indexElement{k}
\def\varm{m}

\def\bitSymbol{b}

\def\numberOfActiveSamples{N_{\rm active}}

\def\complexNumbers{\mathbb{C}}
\def\realNumbers{\mathbb{R}}

\def\numberOfTaps{\mathcal{L}}
\def\tapIndex{l}
\def\decayRate{\tau}

\documentclass[journal]{IEEEtran}
\usepackage[T1]{fontenc}
\usepackage[utf8]{inputenc}
\usepackage{mathtools}
\usepackage{amsthm}
\usepackage{amssymb}
\usepackage{acronym}
\usepackage{hyperref}
\usepackage{stfloats}
\usepackage{xcolor}
\usepackage{amsfonts}
\usepackage{acronym}
\usepackage{upgreek}
\usepackage{cite}
\usepackage{amsmath,amssymb}
\usepackage[english]{babel}
\usepackage[font=footnotesize]{caption}
\usepackage{subcaption}
\usepackage[geometry]{ifsym}
\usepackage{dsfont}
\usepackage{booktabs}
\usepackage{graphicx}

\makeatletter
\newif\ifAC@uppercase@first
\def\Aclp#1{\AC@uppercase@firsttrue\aclp{#1}\AC@uppercase@firstfalse}
\def\AC@aclp#1{%
  \ifcsname fn@#1@PL\endcsname
    \ifAC@uppercase@first
      \expandafter\expandafter\expandafter\MakeUppercase\csname fn@#1@PL\endcsname
    \else
      \csname fn@#1@PL\endcsname
    \fi
  \else
    \AC@acl{#1}s
  \fi 
}
\edef\AC@uppercase@write{\string\ifAC@uppercase@first\string\expandafter\string\MakeUppercase\string\fi\space}
\def\AC@acrodef#1[#2]#3{%
  \@bsphack
  \protected@write\@auxout{}{%
    \string\newacro{#1}[#2]{\AC@uppercase@write #3}%
  }\@esphack
}
\def\Acl#1{\AC@uppercase@firsttrue\acl{#1}\AC@uppercase@firstfalse}
\makeatother

\acrodef{CP}{cyclic prefix}
\acrodef{OOK}{on-off keying}
\acrodef{WUR}{wake-up radio}
\acrodef{WUS}{wake-up signal}
\acrodef{PAPR}{peak-to-average-power ratio}
\acrodef{WFC}{waveform coding}
\acrodef{OFDM}{orthogonal frequency division multiplexing}
\acrodef{MPC}{multi-path channel}
\acrodef{GI}{guard interval}
\acrodef{RMSE}{root-mean-squared error}
\acrodef{IDFT}{inverse discrete Fourier transform}
\acrodef{DFT}{discrete Fourier transform}
\acrodef{ISL}{integrated side lobe}
\acrodef{CAN}{cyclic algorithm-new}
\acrodef{SCAN}{shaping with CAN}
\acrodef{FFT}{fast Fourier transform}
\acrodef{BER}{bit error rate}
\acrodef{AWGN}{additive white Gaussian noise}
\acrodef{QAM}{quadrature amplitude modulation}
\acrodef{SNR}{signal-to-noise ratio}
\acrodef{PDP}{power delay profile}
\acrodef{PPM}{pulse-position modulation}
\acrodef{ACI}{adjacent-channel interference}
\acrodef{PA}{power amplifier}
\acrodef{BLE}{Bluetooth Low Energy}
\acrodef{RF}{radio frequency}
\acrodef{FSK}{frequency-shift keying}
\acrodef{OFDMA}{orthogonal frequency division multiple accessing}
\acrodef{MAC}{medium access control}
\acrodef{WURx}{wake-up radio receiver}
\acrodef{WUTx}{wake-up radio transmitter}
\acrodef{LPF}{low-pass filter}

\begin{document}
\title{ 
Sequence-Based OOK for Orthogonal Multiplexing of Wake-up Radio Signals and OFDM Waveforms
}
\author{Alphan~\c{S}ahin,~\IEEEmembership{Member,~IEEE,}
        and~Rui~Yang,~\IEEEmembership{Member,~IEEE}
\thanks{Dr.~Alphan~\c{S}ahin and Dr.~Rui~Yang are affiliated with InterDigital, Huntington Quadrangle, Melville, NY. email: \{alphan.sahin, rui.yang\}@interdigital.com}}


\maketitle

\begin{abstract}
In this study, we propose an approach to construct \ac{OOK} symbols for \acp{WUR}  by using sequences in frequency domain. The proposed method enables orthogonal multiplexing of \acp{WUS} and \ac{OFDM} waveforms. We optimize the sequences with a tractable algorithm relying on an alternating minimization by considering the reliability of \acp{WUS} in fading channels. We demonstrate the performance of four optimized sequences and compare with state-of-the-art approaches. We show that the proposed method improves the \ac{WUR} performance by controlling the energy distribution in frequency domain while removing the interference-floor at \ac{OFDM} receiver.
\end{abstract}

\acresetall

\section{Introduction}

\acrodef{IoT}{Internet of things}

The functionality of the Internet has been expanded in many dimensions in recent years. The inclusion of the \ac{IoT} with a large number of connected devices in the communication networks significantly changes the role of the Internet in our daily life and, in many situations, the way we interact with the things around us and the way they interact among them.  Combined with other advanced technologies, such as artificial intelligence and cloud-based communication networks, \ac{IoT} enables  broad applications such as smart cars, smart home, smart cities, and smart farms. The root building blocks of those smart things are large number of \ac{IoT} devices, e.g., cameras, actuators, and any small and low-cost electronics which may be wired or connected wirelessly. 

Building the networks with \ac{IoT} devices faces many challenges. As most of the \ac{IoT} devices are the electronics connected remotely  and powered by batteries, minimizing the power consumption to maintain long battery life over multiple years is one of the key design criteria. Since many \ac{IoT} devices only need to operate with very long duty cycle, e.g., once a day or several days for smart meters, or only need to operate when there is a demand triggered by other rarely happened events, e.g., severe weather condition, the concept of \ac{WUR} has been introduced for wireless \ac{IoT} devices.  With \ac{WUR}, an \ac{IoT} device can be put to sleep when it is not used. The \ac{WURx} is a specially designed receiver with extremely low power consumption or without using a battery at all. Such a \ac{WURx} will wake up the \ac{IoT} device once a \ac{WUS} is received.

The concept of WUR has been introduced in several standards in the wireless industry. For example, the working group of IEEE 802.11 has started a project to create a specification amendment \cite{baPAR_2016}, named as 802.11ba, for Wi-Fi devices which may be used for \ac{IoT} applications \cite{McCormick_2017}. 3GPP has also started a discussion to include protocols that enable \ac{WUR} functionalities.  \ac{WUR} should be a simple and low-cost radio without complex signal processing components (e.g., \ac{DFT}) and the corresponding \ac{WUS} may not need to be very sophisticated in some cases. However, the design and generation of such a signal are not straightforward when considering the efficiency of the overall network and the usage of \acp{WUS} in future \ac{IoT} applications. For example, using a dedicated resource (frequency or time) for transmitting \ac{WUS} is inefficient, and combining it with other signals (e.g., \ac{OFDM} signal) without interfering with each other is highly desirable. In this paper, we will explore the ideas of multiplexing of \ac{WUS} with \ac{OOK} symbols and \ac{OFDM} signals for data communication with other devices.

In the literature, several modulation schemes  are proposed for \acp{WUR}. For example, in \cite{kim_2016}, a special bit stream which generates a \ac{WUS} with \ac{OFDM} symbols are proposed by considering IEEE 802.11a/g/n packet structure. One of the proposed bit streams is constructed such that it purposely yields \ac{OFDM} symbols with high \ac{PAPR} and the location of the peak sample is controlled in time to achieve \ac{PPM}. Although the  method has its own merit, in practice, \ac{OFDM} symbols with high \ac{PAPR} may not be reliable. The power back-off due to the high peak sample may limit the coverage range. The distortion due to non-linear hardware can also cause spectral growth which can violate the \ac{ACI} requirements of some standards. Another strategy mentioned in \cite{kim_2016} is to map the low-power constellation points to the lower half of the channel bandwidth, and high-power points to the upper half to encode a bit 0 (and vice versa for bit 1), i.e., \ac{FSK} which is also exploited in several \ac{WURx} designs, e.g., \cite{Im_2017}. In \cite{tang_2015} and \cite{Roberts_2016}, a scheme which modulates the frame length by using multiple consecutive Wi-Fi or \ac{BLE} packets is introduced for \acp{WUR}. The main bottleneck with frame-length modulation is that the data rate can be very low as the \ac{WUS} requires many consecutive packets. In addition, the \ac{BER} performance can degrade drastically as the continuity of the packets for \ac{WUR} can be broken if another station accesses the wireless medium. In \cite{zhang_2018},  \ac{OOK} symbols in a \ac{WUS} are generated through by modulating  \ac{OFDM} symbols via high-power and low-power constellation points for ON and OFF symbols, respectively, for low-data rates. Several other modulation techniques for \acp{WUR} can also be found in \cite{Piyare_2017} and the references therein. To the best of our knowledge,  the design of multiple \ac{OOK} symbols within one \ac{OFDM} symbol duration and the orthogonal multiplexing of \ac{WUS} and \ac{OFDM} waveforms have not been addressed yet.

For IEEE 802.11ba, it has been agreed that the \ac{WUS} will be transmitted by using \ac{OOK} symbols. The \ac{OFDM} waveform parameters adopted in IEEE 802.11n/ac are considered as the baseline for the \ac{OOK} symbol to reuse the existing 802.11n/ac transmitters. The bandwidth of \ac{WUS} is decided to be $4~$MHz and the \ac{OOK} symbol duration is set to $2~\mu$s for high-data rate \ac{WUS}, which is half of the IEEE 802.11n/ac \ac{OFDM} symbol duration. To enable non-coherent detection, the bits for the wake-up payload are encoded with Manchester coding, which implicitly yields \ac{PPM}. Several ways of generating the \ac{OOK} symbols are proposed. For example, using the half of the 802.11n/ac OFDM symbols with $13$ non-zero subcarriers (i.e., masked-based approach)  or using a smaller \ac{IDFT} with $7$ non-zero subcarriers are some of the proposed methods to generate the \ac{OOK} symbols (e.g., see \cite{vinod_2018} and the references in \cite{sahin_2017}). Multiplexing \ac{QAM} and \ac{OOK} symbols in an \ac{OFDMA} framework (such as IEEE 802.11ax) has not been addressed in IEEE 802.11ba. One of the  reasons is because the orthogonality \ac{OFDM} and \ac{OOK} symbols with the state-of-the-art designs cannot be maintained when their symbol durations are different.

In this study, we propose  to construct \ac{OOK} symbols within one \ac{OFDM} symbol for \acp{WUR} by using a fixed number of subcarriers. The proposed approach enables a radio to transmit \acp{WUS} without any interference to the other subcarriers which may potentially be utilized for the data transmission for the other users in the network. We particularly focus on OOK symbols with Manchester coding within an \ac{OFDM} symbol for non-coherent detection. To increase the reliability of \ac{WUS}, we optimize the sequences with a tractable algorithm by taking the leakage on the OFF period and the flatness of the corresponding waveform in time and frequency domain into account. The proposed algorithm is based on an algorithm called \ac{CAN} which was proposed to generate a unimodular sequence with low \ac{ISL} metric \cite{stoica2009}. A sequence generated with \ac{CAN} can lead to significantly low \ac{PAPR} when it is used in \ac{OFDM} due to the low \ac{ISL} of the sequence \cite{Stoica_2009, he2009, soltanalian2013}. Hence, the ideal shape that \ac{CAN} targets is a flat signal in time in terms of sample power. We extend \ac{CAN} to generate any shape in time  and utilize the derived algorithm, i.e., \ac{SCAN}, to achieve orthogonal \ac{OOK} symbols.

The rest of the paper is organized as follows. In Section \ref{sec:systemModel}, we provide system models for the transmitter, channel, and receiver. In Section \ref{sec:sequenceBased}, we discuss the sequences-based \ac{OOK} and provide the details of the proposed algorithm, i.e., \ac{SCAN}. In Section \ref{sec:numerical}, we present numerical results based on IEEE 802.11n/ac waveform parameters. We conclude the paper in Section \ref{sec:conclusion} with final remarks.

{\em Notation:} Hermitian operation is denoted by $(\cdot)^{\rm H}$. The operations $\tilde{(\cdot)}$ and $(\cdot)^*$  reverses the order of the elements and applies element-wise complex conjugation to their arguments, respectively. The operator $\{\cdot \}_\indexElement$ gives the $(\indexElement+1)$th element of its argument,  The 2-norm and infinity norm are denoted by $\lVert{\cdot}\rVert_2$ and $\lVert{\cdot}\rVert_\infty$, respectively. $\expectedValue[\cdot]$ represents the expectation operator. The field of complex numbers is represented by $\complexNumbers$. $\Re\{\cdot\}$ and $\Im\{\cdot\}$ return the real part and the imaginary part of their arguments, respectively.

\section{System Model}
\label{sec:systemModel}
In this section, we provide the models for the \ac{WUTx}, multipath channel, and \ac{WURx}.
\subsection{Wake-up Radio Transmitter}
Consider a communication system where the transmitter generates a \ac{WUS} by mapping the bits to the basis functions in a basis $\waveformBasis$. In this study,
we assume that $\waveformBasis$ consists of two basis functions, i.e., $\waveformBasis=\{\signalTime[{\sequence[\indexSequence]}][\timeVariable]|\indexSequence \in \{0,1\}, \expectedValue[{|\signalTime[\sequence[\indexSequence]][\timeVariable]|^2}] = 1, 0\le\timeVariable<\symbolDuration\}$, and  $\signalTime[{\sequence[0]}][\timeVariable]$ and $\signalTime[\sequence[1]][\timeVariable]$ correspond to the transmitted signals for $\bitSymbol_\varm=0$ and $\bitSymbol_\varm=1$, respectively, where   $ \bitSymbol_\varm$ is the $\varm$th bit  in the \ac{WUS}.  The signal $\signalTime[{\sequence[\indexSequence]}][\timeVariable]$ is constructed by 
 mapping a sequence $\sequence[\indexSequence]\in\complexNumbers^{\lengthOfSequence\times1}$ from frequency domain using $\lengthOfSequence$ contiguous orthogonal subcarriers to time domain as 
\begin{align}
\signalTime[{\sequence[\indexSequence]}][\timeVariable] = 
\frac{1}{\sqrt{\power[]}}\sum_{\indexElement=0}^{\lengthOfSequence-1}\elementOfSequenceMapped[\indexSequence][\indexElement]{\rm e}^{{\rm j}2\pi(\indexElement-\frac{\lengthOfSequence+1}{2})\deltaF\timeVariable}~,
\label{eq:basisFunction}
\end{align}
where  
$\indexElement$ is the subcarrier index,
$\deltaF$ is the subcarrier spacing, 
$||\sequence[\indexSequence]||^2=\power[]$, and
$\symbolDuration=1/\deltaF$. 
Without loss of generality, we assume that $\lengthOfSequence$ is an odd positive integer number, the element of $\sequence[\indexSequence]$ that corresponds to the DC tone is set to zero (i.e., $\elementOfSequenceMapped[\indexSequence][\frac{\lengthOfSequence-1}{2}]=0$), and $\power[]=\lengthOfSequence-1$. The center tone is not utilized because a DC blocker may be employed at the \ac{WUR}. 
Assuming a \ac{CP} is used for the \ac{OFDM} signal, to achieve orthogonal multiplexing of \ac{WUS} and \ac{OFDM} waveforms, a \ac{CP} with the duration of $\cpDuration$ is prepended to $\signalTime[{\sequence[\indexSequence]}][\timeVariable]$.

To enable non-coherent detection  at the \ac{WUR}, we assume that $\signalTime[{\sequence[\indexSequence]}][\timeVariable]$ confines the symbol energy on the period denoted by $\onPeriod[,\indexSequence]= (\indexSequence\Tactive,(\indexSequence+1)\Tactive)$, where $\Tactive\le \symbolDuration/2$ is the active duration for  ON and OFF periods. This choice implicitly corresponds to \ac{PPM} or a  waveform generated via plain \ac{OOK} waveform when bits are encoded with Manchester coding. For example, Manchester coding can result in  $\{0,~1\}$ and $\{1,~0\}$ for $\bitSymbol_\varm = 0$ and $\bitSymbol_\varm = 1$, respectively, and the plain \ac{OOK} symbol duration is set to $\Tactive$ for each coded bit.

We represent the ideal shape for  $|\signalTime[\sequence[\indexSequence]][\timeVariable]|^2$ as
$\shapeTime[\indexSequence][\timeVariable]$ where $\shapeTime[\indexSequence][\timeVariable]=\symbolDuration/\Tactive$ for $\timeVariable \in  \onPeriod[,\indexSequence]$ and $0$ for  $\timeVariable \in  \offPeriod[,\indexSequence]$, where $\offPeriod[,\indexSequence]$ consists of the periods which contain low-energy samples. We assume that $\onPeriod[,\indexSequence]\cap\offPeriod[,\indexSequence]=\emptyset$, $\onPeriod[,\indexSequence]\cup\offPeriod[,\indexSequence]= [0,\symbolDuration)$,  and $\expectedValue[{\shapeTime[\indexSequence][\timeVariable]}] = 1$.

\subsection{Channel Model}
The wireless channel is modeled as an exponential \ac{PDP} with $\numberOfTaps$ independent taps where the unnormalized power of the $\tapIndex$th tap is expressed as ${\rm e}^{-\decayRate\tapIndex}$ and $\decayRate$ corresponds to the decay rate. Note that $\decayRate=0$ yields a uniform \ac{PDP}. We assume  Rayleigh distribution for the amplitude of each tap.

\def\cutoffFrequency{f_{\rm c}}
\subsection{Wake-up Radio Receiver}
 We assume that \ac{WUR} skips the samples during \ac{CP} durations and detects $\bitSymbol_\varm$ by comparing the energy on active durations without any channel estimation. We also consider that  that \ac{WUR} employs a second order Butterworth \ac{LPF} where its cut-off frequency is denoted by $\cutoffFrequency$. 

%
\section{Sequence-Based On-Off Keying}
\label{sec:sequenceBased}
In this section, we discuss sequence-based \ac{OOK} and the corresponding algorithm, i.e., \ac{SCAN}, to obtain the sequences.
\subsection{Problem Formulation}
In this study, our main goal is to find the sequence $\sequence[\indexSequence]$ such that the Euclidean distance between $|\signalTime[\sequence[\indexSequence]][\timeVariable]|^2$ and $\shapeTime[\indexSequence][\timeVariable]$ is minimum while satisfying certain constraints related to the leakage during the OFF period and flatness in time and frequency.

In the sequel, we only consider the optimization of $\sequence[0]$ and omit the index $\indexSequence$ unless otherwise stated 
since $\signalTime[\sequence[0]][\timeVariable]$ can be generated through $\signalTime[\sequence[1]][\timeVariable]$. 
For example, if $\Tactive$ is chosen as $\symbolDuration/2$, $|\signalTime[\sequence[0]][\timeVariable]|$ can be the time-reversal of $|\signalTime[\sequence[1]][\timeVariable]|$. One way of achieving the time-reversal relationship between the basis functions is to relate $\sequence[1]$ and $\sequence[0]$ as $\sequence[1]=\sequenceFlipped[0]$ or  $\sequence[1]=\sequenceConjugated[0]$  since the inverse Fourier transformation of $\sequenceFlipped[0]$ and $\sequence[1]=\sequenceConjugated[0]$ are $\signalTime[\sequence[0]][-\timeVariable]$ and $\signalTimeConjugated[\sequence[0]][-\timeVariable]$, respectively. The choice of $\sequence[1]=\sequenceFlipped[0]$ reduces the hardware complexity at the transmitter since the transmitter can generate the basis functions by changing the order of the elements of $\sequence[0]$. In the cases where $\Tactive$ is less than $\symbolDuration/2$, $\{\sequence[1]\}_{\indexElement}$ can be set to  ${\rm e}^{-{\rm j}2\pi\indexElement\Tactive/\symbolDuration}\{\sequence[0]\}_{\indexElement}$ by exploiting the time-shifting property of the inverse Fourier transformation. 

We express the distance between $|\signalTime[\sequence[]][\timeVariable]|^2$ and $\shapeTime[][\timeVariable]$ with \ac{RMSE}  as 
\begin{align}
\cost[] = \expectedValue[{\left||\signalTime[\sequence[]][\timeVariable]|^2 - \power[]\shapeTime[][\timeVariable] \right|^2}]^{\frac{1}{2}}~.
\label{eq:costFunction}
\end{align}
The optimization problem that  gives the optimal sequence for $\signalTime[\sequence[]][\timeVariable]$  can then be written as
\begin{align}
\sequenceOptimal[] = \arg \min_{\sequence[]} & \expectedValue[{\left||\signalTime[\sequence[]][\timeVariable]|^2 - \power[]\shapeTime[][\timeVariable] \right|^2}]^{\frac{1}{2}}+\lagrangeVariable\lVert{\sequence[]}\rVert_{\infty}
\label{eq:optimizationProblemMain}
\\  \text{s.t.} 
& ~~  \text{c$_1$:~}  \elementOfSequenceMapped[][\frac{\lengthOfSequence-1}{2}]=0  \nonumber \\
& ~~  \text{c$_2$:~} |\signalTime[\sequence[]][0]|^2 \le \powerFirstSample  \nonumber\\
& ~~  \text{c$_3$:~}  |\signalTime[\sequence[]][\timeVariable]|^2\le \powerLeakage \nonumber, 
\timeVariable\in\offPeriod[] \nonumber 
~,
\end{align}
where  $\powerFirstSample$ and  $\powerLeakage$ are the desired values for the basis functions. In \eqref{eq:optimizationProblemMain}, we omit the frequency-shift component in \eqref{eq:basisFunction} as it does not change the cost given in \eqref{eq:costFunction}. 
We also regularize the elements of $\sequenceOptimal[]$ by introducing a regularization term $\lVert{\sequence[]}\rVert_{\infty}$  to the \ac{RMSE} with a  parameter denoted by $\lagrangeVariable\ge 0$. When $\lagrangeVariable> 0$, the regularization term in \eqref{eq:optimizationProblemMain} applies a penalty if the sequence energy is localized on few elements and increase the reliability of the waveform against frequency-selective channels. 
The constraints c$_1$, c$_2$, and c$_3$ aim at zero DC tone, smooth ramp-up for the ON period, and low instantaneous power on the OFF period, respectively.  Unfortunately, the argument of the expectation operator in \eqref{eq:optimizationProblemMain} is a quartic function. Therefore, \eqref{eq:optimizationProblemMain} is not a convex optimization problem. In the following part, we propose a tractable algorithm to solve \eqref{eq:optimizationProblemMain}. 

\subsection{SCAN: Shaping with CAN}
\label{sec:SCAN}
To calculate the expected value in \eqref{eq:optimizationProblemMain}, we sample \eqref{eq:costFunction} within the period of $[0,\symbolDuration)$, where the $\indexSample$th sampling time is given by $\indexSample/\numberOfSamples\times\symbolDuration$, $\indexSample\in\{0,1,\dots,\numberOfSamples-1\}$,  and $\numberOfSamples$ is the number of samples. The \ac{RMSE} in \eqref{eq:costFunction} can then be  calculated approximately as
\begin{align}
\costSampled[] = \left(\frac{1}{\numberOfSamples}\sum_{\indexSample=0}^{\numberOfSamples-1} {\left|\left|\sum_{\indexElement=0}^{\lengthOfSequence-1} \elementOfSequenceMapped[][\indexElement]{\rm e}^{{\rm j}2\pi\indexElement\frac{\indexSample}{\numberOfSamples}}\right|^2 - \power[]\shapeTime[][\frac{\indexSample\symbolDuration}{\numberOfSamples}] \right|^2}\right)^{\frac{1}{2}}~.
\label{eq:costSampled}
\end{align}
Due to the convolution theorem, the maximum support of the \ac{IDFT} of the  $|\signalTime[\sequence[]][\indexSample\symbolDuration/\numberOfSamples]|^2$ is $2\lengthOfSequence-1$ when $\numberOfSamples\ge 2\lengthOfSequence-1$. Hence, $\numberOfSamples\ge 2\lengthOfSequence-1$ in \ref{eq:costSampled} must also hold true to oversample the basis function $|\signalTime[\sequence[]][\timeVariable]|^2$.

To enable tractable algorithms, we propose to modify \eqref{eq:costSampled}  as
\begin{align}
\costSampledModified[] =& \left(\frac{1}{\numberOfSamples}\sum_{\indexSample=0}^{\numberOfSamples-1} {\left|\sum_{\indexElement=0}^{\lengthOfSequence-1} \elementOfSequenceMapped[][\indexElement]{\rm e}^{{\rm j}2\pi\indexElement\frac{\indexSample}{\numberOfSamples}} -\sqrt{ \power[]}\elementOfAmplitudeShape[][\indexSample]{\elementOfPhaseShape[][\indexSample]} \right|^2} \right)^{\frac{1}{2}}\nonumber
\\
=&\frac{1}{\sqrt{\numberOfSamples}}||\DFTmatrix[\numberOfSamples]^{\rm H}\mappingMatrix\sequence[] -\sqrt{\power[]}\amplitudeMatrix\phaseShape[]||_2~,
\label{eq:newCost}
\end{align}
where
$\elementOfAmplitudeShape[][\indexSample] = \sqrt{\shapeTime[][\indexSample\symbolDuration/\numberOfSamples]}$, 
$\DFTmatrix[\numberOfSamples]\in\complexNumbers^{\numberOfSamples\times\numberOfSamples}$ is the \ac{DFT} matrix, 
$\mappingMatrix\in\realNumbers^{\numberOfSamples\times\lengthOfSequence}$ is the mapping matrix, 
$\amplitudeMatrix\in\realNumbers^{\numberOfSamples\times\numberOfActiveSamples}$ is a semi-orthogonal matrix and obtained removing the zero columns of $\diag[{\amplitudeShape[]}]$, 
$\numberOfActiveSamples$ is the number of non-zero elements in vector $\amplitudeShape[]$, and
$\phaseShape[]\in\complexNumbers^{\numberOfActiveSamples\times 1}$ is a phase vector.

Both \eqref{eq:costSampled} and \eqref{eq:newCost} are metrics to measure the distance  between $|\signalTime[\sequence[]][\timeVariable]|^2$ and $\shapeTime[][\timeVariable]$, i.e., if $\costSampledModified[]$ is small, $\costSampled[]$ also becomes small. However, \eqref{eq:newCost} achieves it with a quadratic function, i.e., reduces the order of \eqref{eq:costSampled}. The function in \eqref{eq:newCost} can be thought as a generalization of the approach proposed in \cite{stoica2009}. The main objective in \cite{stoica2009} is to minimize the \ac{ISL} metric of a unimodular sequence $\sequence[]$, i.e., equivalent to \eqref{eq:costSampled} for $\shapeTime[][\timeVariable] = 1$, $|\elementOfSequenceMapped[][\indexElement]|=1$, and $\power[]=\lengthOfSequence$. An iterative algorithm based on \ac{FFT}, called \ac{CAN}, is derived by replacing the cost function in \eqref{eq:costSampled} with an another cost function, i.e., equivalent to a special case of \eqref{eq:newCost}.  The discussion related to the validity of the modification is provided in \cite{he2009}.
Since the \ac{CAN} minimizes the \ac{ISL} metric of the sequence, it also targets the sequences that lead to low \ac{PAPR} when the same sequence is used in frequency with an \ac{OFDM} waveform, i.e., ultraflat polynomials \cite{Erdelyi01polynomialswith, erdos1957}. In this study, the proposed algorithm is based on \ac{CAN}, called \ac{SCAN}, and aim at the optimal sequences that lead to {\em any} ideal signal shape in time.

We first rewrite the optimization problem given in \eqref{eq:optimizationProblemMain} by using the cost function given in \eqref{eq:newCost}  as
\begin{align}
\sequenceOptimal[] = \arg \min_{\sequence[],\phaseShape[]} &\frac{1}{\sqrt{\numberOfSamples}}||\DFTmatrix[\numberOfSamples]^{\rm H}\mappingMatrix\sequence[] -\sqrt{ \power[]}\amplitudeMatrix\phaseShape[]||_2+ \lagrangeVariable\lVert{\sequence[]}\rVert_{\infty}
\label{eq:optimizationProblemMainNoAlphabetMatrix}
\\  \text{s.t.} 
& ~~  \text{c$_1$:~}  \elementOfSequenceMapped[][\frac{\lengthOfSequence-1}{2}]=0 \nonumber
\\  & ~~  \text{c$_2$:~}  |\{\DFTmatrix[\numberOfSamples]^{\rm H}\mappingMatrix\sequence[]\}_{0}|^2 \le  \powerFirstSample \nonumber
\\  & ~~  \text{c$_3$:~}  |\{\DFTmatrix[\numberOfSamples]^{\rm H}\mappingMatrix\sequence[]\}_{\indexSample}|^2 \le \powerLeakage, \indexSample \in \offPeriodSampled[] \nonumber
\\  & ~~  \text{c$_4$:~} |\{\phaseShape[]\}_\indexSample|^2=1, \indexSample \in  \nonumber \onPeriodSampled[]
~,
\end{align}
where  $\lagrangeVariable$ is a Lagrange variable,  $\onPeriodSampled[] = \{\indexSample|\elementOfAmplitudeShape[][\indexSample]\neq 0\}$ and $\offPeriodSampled[] = \{\indexSample|\elementOfAmplitudeShape[][\indexSample] = 0\}$. The problem in \eqref{eq:optimizationProblemMainNoAlphabetMatrix} is still non-convex due to the constraint  c$_4$. However, the local optima can still be obtained by splitting \eqref{eq:optimizationProblemMainNoAlphabetMatrix} into two subproblems and solving them iteratively, i.e., alternating minimization. 

For the $\indexIteration$th iteration, the first subproblem for a given $\sequence[]^{(\indexIteration-1)}$ is given by
\begin{align}
\phaseShape[]^{(\indexIteration)} = \arg \min_{\phaseShape[]} &||\DFTmatrix[\numberOfSamples]^{\rm H}\mappingMatrix\sequence[]^{(\indexIteration-1)} -\sqrt{ \power[]}\amplitudeMatrix\phaseShape[]||_2
\label{eq:optimizationProblemMainNoAlphabetMatrixProblemA}
\\  \text{s.t.} &~~ |\{\phaseShape[]\}_\indexSample|^2=1, \indexSample \in  \nonumber \onPeriodSampled[]
~.
\end{align}
Although \eqref{eq:optimizationProblemMainNoAlphabetMatrixProblemA} is not a convex problem due to the unimodular vector constraint, the global optimum can be obtained via a closed-form solution given by $\{\phaseShape[]\}_\indexSample = \angle \{\DFTmatrix[\numberOfSamples]\mappingMatrix\sequence[] \}_\indexSample $. This is due to the fact that the support of the columns of $\amplitudeMatrix$  do not intersect (i.e., semi-orthogonal matrix) and the non-zero elements are positive (i.e., does not change the phase of $ \{\DFTmatrix[\numberOfSamples]\mappingMatrix\sequence[] \}_\indexSample $). 

The second subproblem can be expressed as
\begin{align}
\sequence[]^{(\indexIteration)} &= \arg  \min_{\sequence[]}\frac{1}{\sqrt{\numberOfSamples}} ||\DFTmatrix[\numberOfSamples]^{\rm H}\mappingMatrix\sequence[] -\sqrt{ \power[]}\amplitudeMatrix\phaseShape[]^{(\indexIteration)}||_2 + \lagrangeVariable\lVert{\sequence[]}\rVert_{\infty}
\label{eq:optimizationProblemMainNoAlphabetMatrixProblemB}
\\  \text{s.t.} &~~  \text{c$_1$:~}  \elementOfSequenceMapped[][\frac{\lengthOfSequence-1}{2}]=0 \nonumber
 \\  & ~~\text{c$_2$:~}   \Re\{\{\DFTmatrix[\numberOfSamples]^{\rm H}\mappingMatrix\sequence[\indexSequence]\}_{0}\}, \Im\{\{\DFTmatrix[\numberOfSamples]^{\rm H}\mappingMatrix\sequence[]\}_{0}\}\le \boundFirst \nonumber
 \\  &  ~~\text{c$_3$:~}   \Re\{\{\DFTmatrix[\numberOfSamples]^{\rm H}\mappingMatrix\sequence[]\}_{\indexSample}\}, \Im\{\{\DFTmatrix[\numberOfSamples]^{\rm H}\mappingMatrix\sequence[]\}_{\indexSample}\}\le \boundLeakage,  \indexSample \in  \nonumber \offPeriodSampled[]
~.
\end{align}
In \eqref{eq:optimizationProblemMainNoAlphabetMatrixProblemB}, the quadratic constraints \text{c$_2$} and \text{c$_3$} in \eqref{eq:optimizationProblemMainNoAlphabetMatrix} are replaced with the linear constraints which limit the imaginer and real components of the samples instead of their powers to simplify the problem. Since \eqref{eq:optimizationProblemMainNoAlphabetMatrixProblemB} is a convex optimization problem, it can be solved by using solvers such as SeDuMi \cite{S98guide}.  

The cost function in \eqref{eq:optimizationProblemMainNoAlphabetMatrix} always decreases when \eqref{eq:optimizationProblemMainNoAlphabetMatrixProblemA} and \eqref{eq:optimizationProblemMainNoAlphabetMatrixProblemB} are solved iteratively since the global optima for both \eqref{eq:optimizationProblemMainNoAlphabetMatrixProblemA} and \eqref{eq:optimizationProblemMainNoAlphabetMatrixProblemB} can be obtained. After certain termination condition is met, the optimized sequence $\sequenceOptimalIterations$ is obtained as $\sequenceOptimalIterations = \sqrt{\power[]}\times\sequence[]^{(\indexIteration)}/||\sequence[]^{(\indexIteration)}||_2 $.




\section{Numerical Results}
\label{sec:numerical}
 In this section, we numerically evaluate the sequences generated through the method derived in Section \ref{sec:SCAN}. For the evaluations, we adopt the IEEE 802.11n/ac \ac{OFDM} parameters, i.e., $\symbolDuration = 3.2~\mu$s, $\deltaF = 312.5$ kHz, and $\cpDuration = 0.8~\mu$s. We generate the basis functions in $\waveformBasis$ by using $\lengthOfSequence = 15$ adjacent tones. We consider four different optimized sequences by using the parameter sets given by
$\{ \boundLeakage=\boundFirst=\text{1e-3}, \lagrangeVariable=0,~\Tactive=1.2~\mu\text{s}\}$, $\{ \boundLeakage=\boundFirst=\text{1e-3},~\lagrangeVariable =2.2,~\Tactive = 1.2~\mu\text{s}\}$, $\{ \boundLeakage=\boundFirst=\text{1e-4},~\lagrangeVariable =0,~\Tactive = 1.6~\mu\text{s}\}$, and $\{ \boundLeakage=\boundFirst=\text{1e-4},~\lagrangeVariable = 2.2,~\Tactive = 1.6~\mu\text{s}\}$. For the optimization, the initial sequence $\sequence[]^{(0)}$ is chosen as an all-ones vector of length $15$. The optimized sequences for $\bitSymbol_\varm=0$ are listed in \tablename~\ref{tab:sequences}. In the simulations, $\signalTime[{\sequence[1]}][\timeVariable]$ is generated by circularly shifting $\signalTime[{\sequence[0]}][\timeVariable]$ in time by $\Tactive$.

\begin{table}[!t]
	\centering
	\caption{Sequences}
	\begin{tabular}{l l l l }
		Sequence \#1 & 	Sequence \#2 & 	Sequence \#3 & 	Sequence \#4 \\
				\toprule
     $ 0.117 \angle  22.5^\circ  $  &  $ 0.692 \angle 266.19^\circ  $  &  $ 0.031 \angle 180 ^\circ  $  &  $ 0.225 \angle  50.42 ^\circ  $  \\ 
     $ 0.375 \angle 315.0^\circ  $  &  $ 0.986 \angle 211.15^\circ  $  &  $ 0.171 \angle  90 ^\circ  $  &  $ 0.705 \angle 324.83 ^\circ  $  \\ 
     $ 0.784 \angle 247.5^\circ  $  &  $ 1.119 \angle 157.37^\circ  $  &  $ 0.509 \angle 0^\circ  $  &  $ 1.175 \angle 242.64 ^\circ  $  \\ 
     $ 1.221 \angle 180.0^\circ  $  &  $ 1.119 \angle 115.63^\circ  $  &  $ 1.023 \angle 270 ^\circ  $  &  $ 1.225 \angle 170.47 ^\circ  $  \\ 
     $ 1.480 \angle 112.5^\circ  $  &  $ 1.119 \angle  80.70^\circ  $  &  $ 1.489 \angle 180 ^\circ  $  &  $ 1.052 \angle 122.68 ^\circ  $  \\ 
     $ 1.367 \angle  45.0^\circ  $  &  $ 1.119 \angle  34.18^\circ  $  &  $ 1.557 \angle  90 ^\circ  $  &  $ 1.225 \angle  74.16 ^\circ  $  \\ 
     $ 0.826 \angle 337.5^\circ  $  &  $ 0.735 \angle 337.50^\circ  $  &  $ 1.011 \angle   0 ^\circ  $  &  $ 0.980 \angle      0 ^\circ  $  \\ 
     $ 0     \angle     0^\circ  $  &  $ 0 \angle          0^\circ  $  &  $     0 \angle   0^\circ   $  &  $     0 \angle      0 ^\circ  $  \\ 
     $ 0.826 \angle  22.5^\circ  $  &  $ 0.735 \angle  22.50^\circ  $  &  $ 1.011 \angle   0 ^\circ  $  &  $ 0.980 \angle      0 ^\circ  $  \\ 
     $ 1.367 \angle 315.0^\circ  $  &  $ 1.119 \angle 304.18^\circ  $  &  $ 1.557 \angle 270 ^\circ  $  &  $ 1.225 \angle 254.16 ^\circ  $  \\ 
     $ 1.480 \angle 247.5^\circ  $  &  $ 1.119 \angle 215.70^\circ  $  &  $ 1.489 \angle 180 ^\circ  $  &  $ 1.052 \angle 122.68 ^\circ  $  \\ 
     $ 1.221 \angle 180.0^\circ  $  &  $ 1.119 \angle 115.63^\circ  $  &  $ 1.023 \angle  90 ^\circ  $  &  $ 1.225 \angle 350.47 ^\circ  $  \\ 
     $ 0.784 \angle 112.5^\circ  $  &  $ 1.119 \angle  22.37^\circ  $  &  $ 0.509 \angle   0 ^\circ  $  &  $ 1.175 \angle 242.64 ^\circ  $  \\ 
     $ 0.375 \angle  45.0^\circ  $  &  $ 0.986 \angle 301.15^\circ  $  &  $ 0.171 \angle 270 ^\circ  $  &  $ 0.705 \angle 144.83 ^\circ  $  \\ 
     $ 0.117 \angle 337.5^\circ  $  &  $ 0.692 \angle 221.19^\circ  $  &  $ 0.031 \angle 180 ^\circ  $  &  $ 0.225 \angle  50.42 ^\circ  $  \\ 
		\midrule
		\bottomrule
	\end{tabular}%
	\label{tab:sequences}%
\end{table}

\subsection{Time and Frequency Characteristics}
\begin{figure}[!t]
	\centering	
    {\includegraphics[width =3.5in]{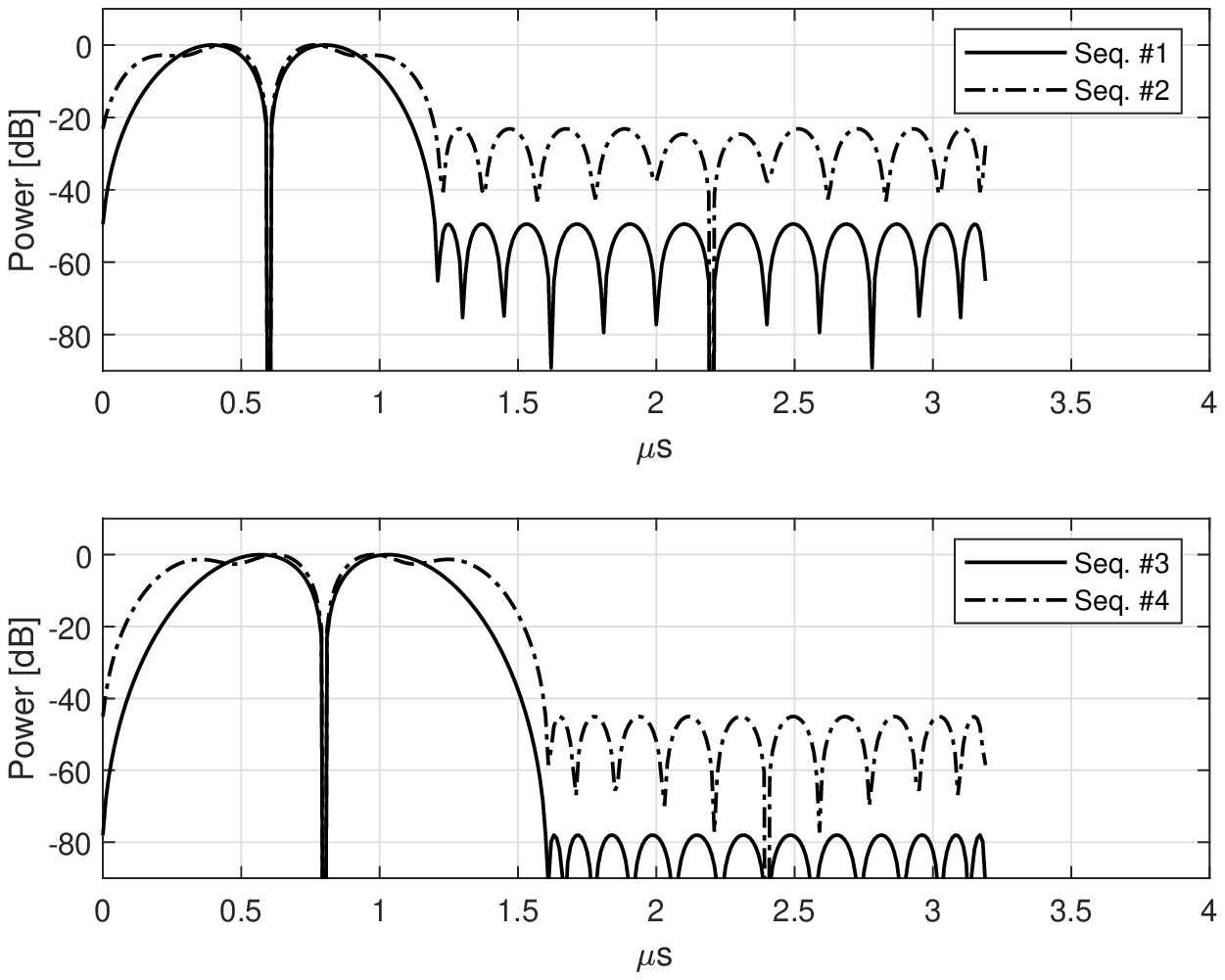}}
	\caption{The transmitted waveform when $\bitSymbol_\varm=0$, i.e., $\signalTime[{\sequence[0]}][\timeVariable]$.}
	\label{fig:basisFuntionsInTime}
\end{figure}

The corresponding waveforms for the sequences in \tablename~\ref{tab:sequences} are provided in \figurename~\ref{fig:basisFuntionsInTime}. While the first and the second sequences confine the energy on the duration of $1.2~\mu$s, the third and the fourth sequences keep the energy on the first half of $\symbolDuration$, i.e., $1.6~\mu$s. Therefore, for the same amount of bit energy, the first and the second sequences cause more peaky signals in time as compared to the third and the fourth sequences. On the other hand,  the first and the second sequences allow smoother transitions between adjacent symbols in \ac{WUS} since the \ac{CP} portion of the corresponding waveforms do not contain high energy samples. This is due to the fact that there are no high-energy samples on the last $0.8~\mu$s of the corresponding waveforms for $\bitSymbol_\varm=0$ and $\bitSymbol_\varm=1$. It is also worth noting that these sequences are  compatible with the \acp{WUR} which cannot discard the \ac{CP} duration since the energy is localized either on the first or the last $2~\mu$s. In addition, They can be utilized for the waveforms which generate an internal guard interval within \ac{IDFT} duration (e.g., \cite{Berardinelli_2013}). 

\begin{figure}[!t]
	\centering
	{\includegraphics[width =3.5in]{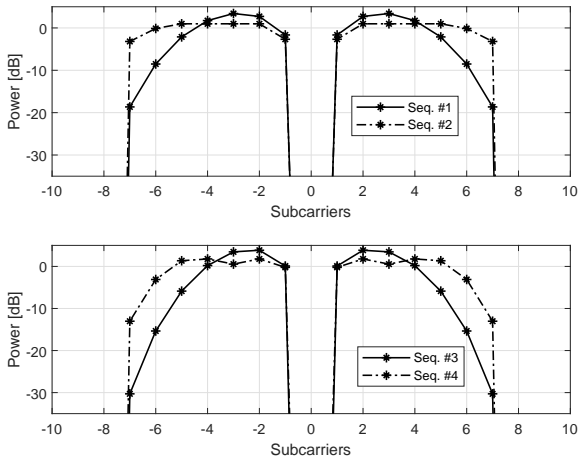}
	}
	\caption{Power distribution of the sequences over the subcarriers.}
	\label{fig:basisFuntionsInFrequency}
\end{figure}

When $\Tactive$ is reduced to $1.6~\mu$s to $1.2~\mu$s, the leakage on the OFF duration increases. As shown in  \figurename~\ref{fig:basisFuntionsInTime}, the ratio between the peak values on the ON and OFF durations is approximately $78$  dB for the third  sequence whereas it reduces to $49.5$ dB  for the first sequence. This reduction is due to the fact that $\offPeriod[,\indexSequence]$ becomes a larger set while $\lengthOfSequence$ is fixed. When $\lagrangeVariable$ is $0$, there is no penalty applied for the non-uniform power distribution of the sequence elements. When $\lagrangeVariable$ is set to $2.2$ to control the power distribution, the difference between the maximum levels on the ON and OFF durations reduces further to $23.2$ dB and $45$ dB for the second and the fourth sequences, respectively. The power distribution over the subcarriers for each sequence are given in \figurename~\ref{fig:basisFuntionsInFrequency}. The first and the third sequences allocate more energy on 6-8 tones while the second and the fourth sequences have more uniform  power distribution across the 15 tones at the expense of more leakage in time.

\subsection{BER Performance for Standalone Scenarios}
\begin{figure}[!t]
	\centering
	{\includegraphics[width =3.5in]{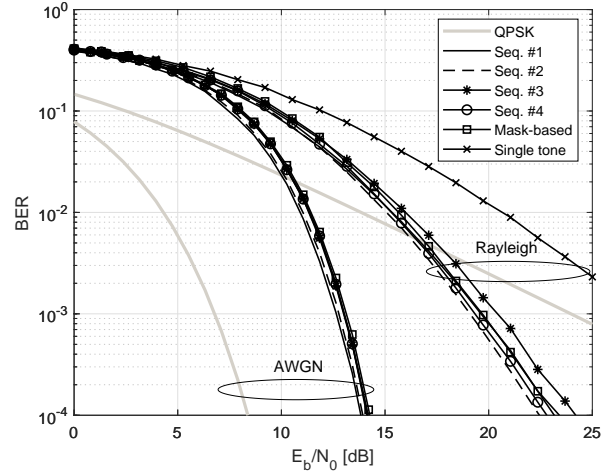}
	}
	\caption{Standalone \ac{BER} performance.}
	\label{fig:standalone}
\end{figure}

In \figurename~\ref{fig:standalone}, we evaluate the \ac{BER} performance when the proposed sequences and the other available \ac{OOK} waveform options, i.e., turning on and off the carrier (single tone) and mask-based approach proposed in \cite{vinod_2018},  are utilized. For multipath channel, we assume that $\numberOfTaps = 10$ and $\decayRate = 0.1$ at the sample rate of $20~$MHz.   We set the sample rate of the \ac{WUR} to $20$ MHz. Note that the sample rate of the \ac{WUR} can be decreased  to reduce the power consumption of the \ac{WUR}. In this case, the corresponding \ac{BER} curve for \ac{WUR} in \figurename~\ref{fig:standalone} needs to be shifted to the right proportionally. We assume that the cut-off frequency $\cutoffFrequency$ of \ac{LPF} at \ac{WUR} is $2.5~$MHz. As shown in \figurename~\ref{fig:standalone},
all of the \ac{OOK} waveforms are at least 6 dB worse than the \ac{BER} performance of the QPSK symbols in the \ac{AWGN} channel. This is due to the fact that the Euclidean distance between the bits is $1/\sqrt{2}$ times that of the QPSK symbols and the non-coherent detection is used at the \ac{WUR}. In AWGN, the \ac{BER} performances for different \ac{OOK} options are similar to each other. The sequence options which cause shorter $\Tactive$ duration perform slightly better as compared to the ones with larger $\Tactive$ duration. This improvement is because the \ac{WUR} accumulates less noise energy when $\Tactive$ is small. The second sequence is slightly worse than the first sequence due to its higher leakage on the OFF duration. Nevertheless, because of less noise accumulation, the second sequence perform better than the third and the fourth sequences although it has a higher leakage at the OFF period. In \ac{AWGN}, single tone, mask-based method, the third sequence, and the fourth sequences performs almost identical. However, in fading channel, the performance of single tone  dramatically deteriorates as it does not exploit the frequency selectivity. The other options are within the margin of 1.5 dB. The best performance is achieved by the second sequence since the power is  well-distributed in frequency as compared to others and a shorter $\Tactive$ is adopted. The third sequence is the worst one as the energy of the sequence is localized in frequency as shown in \figurename~\ref{fig:basisFuntionsInFrequency} and large $\Tactive$ causes more noise accumulation.

\subsection{BER Performance under Multiplexing}
\begin{figure}[!t]
    \centering
    \begin{subfigure}{0.5\textwidth}
        \centering
        \includegraphics[width =3.5in]{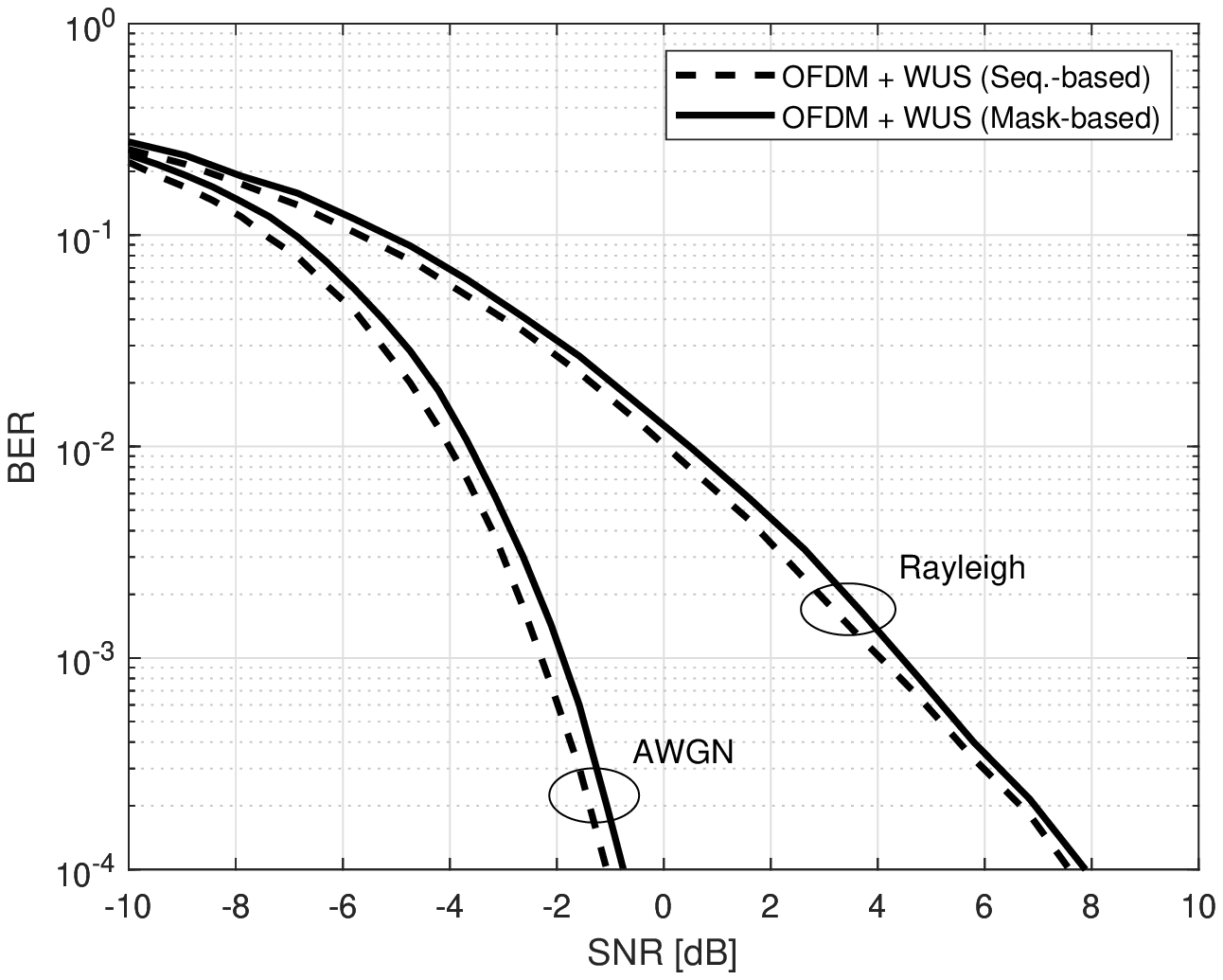}
        \caption{BER at \ac{WUR}}
        \label{fig:WURrx}
    \end{subfigure}\\
    \begin{subfigure}{0.5\textwidth}
        \centering
        \includegraphics[width =3.5in]{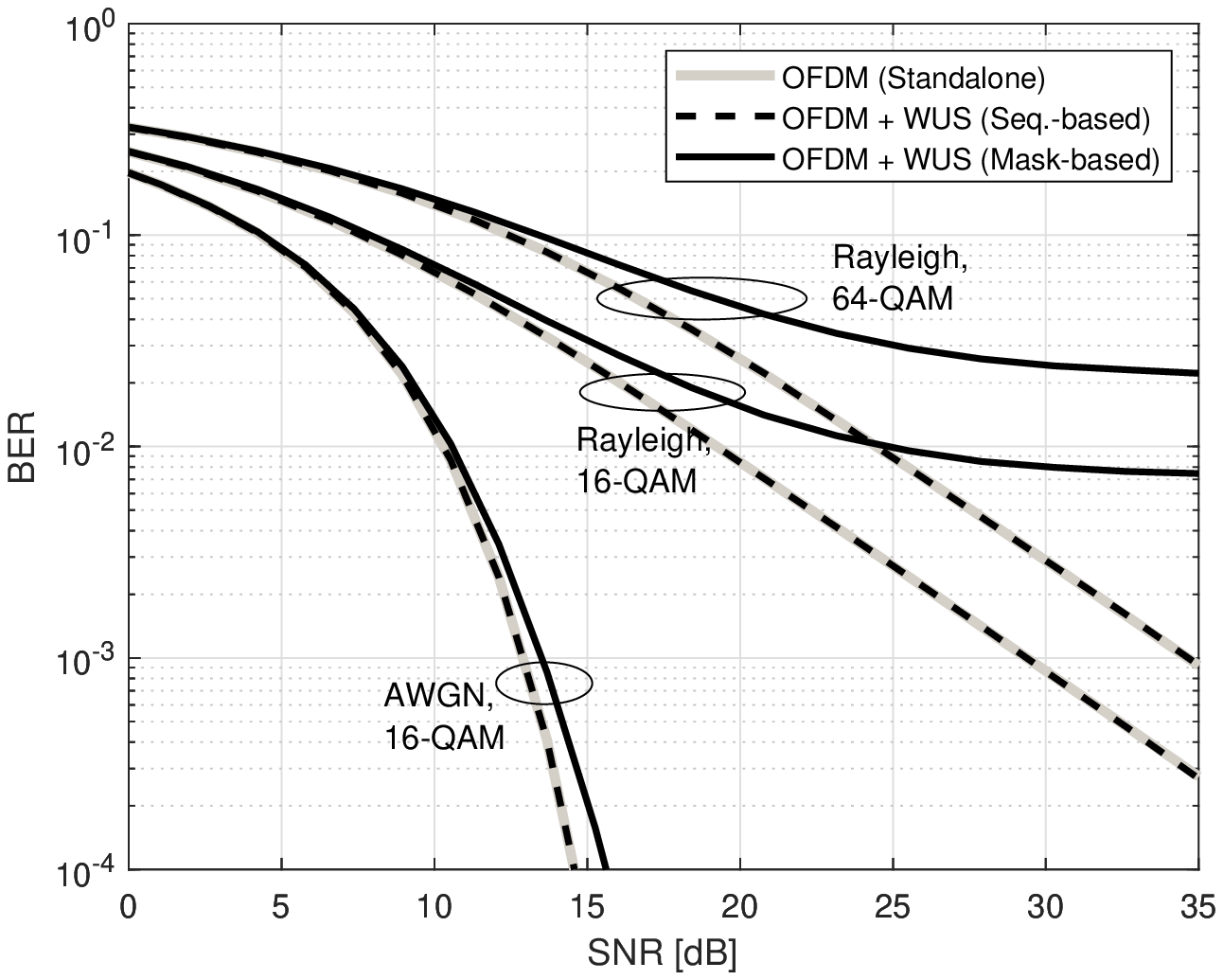}
        \caption{BER at OFDM receiver}
        \label{fig:OFDMrx}
    \end{subfigure}
    \caption{BER performance when \ac{WUS} and QAM symbols are multiplexed in frequency}
    \label{fig:coexsistence}
\end{figure}

In \figurename~\ref{fig:coexsistence}, we evaluate the \ac{BER} performance of the \ac{OFDM} and \ac{WUR} receivers when \ac{WUS} and \ac{QAM} symbols are multiplexed in frequency. We assume that \ac{WUS} is scheduled to the center of the band, i.e., $\{-7,\dots,7\}$ tones. The \ac{QAM} symbols are allocated to the adjacent subcarriers where the tone indices are $\{ -16,\dots,-10,10, \dots, 16\}$. We assume that signal power for \ac{WUS} and \ac{OFDM} symbol are identical and half of the signal power is distributed to 14 \ac{QAM} symbols. The \ac{SNR} is calculated at $20~$MHz sample rate. We evaluate two waveform options for \ac{WUS}, i.e., the proposed scheme with the second sequence and the mask-based method proposed in  \cite{vinod_2018}. In \figurename~\ref{fig:coexsistence}\subref{fig:WURrx}, the \ac{BER} performance at the \ac{WUR} is provided. The \ac{LPF} is able to filter out the QAM symbols on adjacent tones and the \ac{BER} performance at \ac{WUR} for both approaches acceptable. The order of the \ac{BER} curves follows the same order in \figurename~\ref{fig:standalone}. On the other hand, the OFDM receiver is sensitive to the existence of \ac{WUS}. As shown in  \figurename~\ref{fig:coexsistence}\subref{fig:OFDMrx}, the \ac{BER} performance at the \ac{OFDM} receiver degrades drastically when  mask-based \ac{OOK} is utilized for \ac{WUS}. The degradation is because the mask-based \ac{OOK} symbols are not orthogonal to the OFDM symbol and contaminate the adjacent tones which are utilized by \ac{QAM} symbols. On the other hands, the \ac{WUS} with the proposed scheme is always orthogonal to the subcarriers used for \ac{QAM} symbols and do not degrade the performance at the \ac{OFDM} receiver since the \ac{WUS} waveform is generated within the \ac{OFDM} framework.

\section{Concluding Remarks}
\label{sec:conclusion}
In this study, we propose a method which generates \ac{OOK} symbols by using sequences in the frequency domain. 
The proposed method does not cause any interference to the adjacent subcarriers and enables orthogonal multiplexing of \acp{WUS} and \ac{OFDM} waveforms.  
We derive the sequences with a tractable algorithm relying on \ac{CAN} and demonstrate the performance of four  optimized sequences by considering the leakage on the OFF period and the flatness of the corresponding waveforms in time and frequency. The numerical results show that the proposed approach  removes the interference-floor at \ac{OFDM} receiver when \ac{OFDM}-based waveforms and \ac{WUS} are multiplexed in the frequency domain. In addition, the \ac{WUR} performance is improved when the energy distribution of a coded \ac{OOK} symbol in frequency is even and the leakage on the OFF period is low. Multiplexing multiple \acp{WUS}, obtaining scalable expressions of the sequences, and packing more than two \ac{OOK} symbols within one \ac{OFDM} symbol duration are several directions for the extension of this work.

\bibliographystyle{IEEEtran}
\bibliography{sequenceOOK}

\end{document}